\documentclass[12pt]{article}

\usepackage{enumerate}
\usepackage{multirow,graphicx}
\usepackage{amsmath, amsfonts, amssymb, mathrsfs, rotating, setspace}
\usepackage[utf8]{inputenc}
\usepackage{algorithm2e}
\usepackage{hyperref}
\usepackage{url}
\usepackage{natbib}
\usepackage{tikz}
\usepackage{soul}

\newcommand{\blind}{1}

\usepackage{xcolor} 

\usetikzlibrary{calc}
\usetikzlibrary{positioning}
\usetikzlibrary{arrows}

\usepackage{amsthm}

\definecolor{rightcol}{HTML}{FFFF00}
\definecolor{leftcol}{HTML}{0000FF}

\begin{document}

\def\spacingset#1{\renewcommand{\baselinestretch}%
{#1}\small\normalsize} \spacingset{1}


\if1\blind
{
  \title{\bf Sharp nonparametric bounds for decomposition effects with two binary mediators}
  \author{Erin E Gabriel\hspace{.2cm}\\
  and \\
  Michael C Sachs \\
  and \\
  Arvid Sjölander \\
    Department of Medical Epidemiology and Biostatistics \\ Karolinska Institutet, \\
Stockholm,
Sweden.}
  \maketitle
} \fi

\if0\blind
{
  \bigskip
  \bigskip
  \bigskip
  \begin{center}
    {\LARGE\bf Sharp nonparametric bounds for decomposition effects with two binary mediators}
\end{center}
  \medskip
} \fi

\bigskip
\begin{abstract}
In randomized trials, once the total effect of the intervention has been estimated, it is often of interest to explore mechanistic effects through mediators along the causal pathway between the randomized treatment and the outcome. In the setting with two sequential mediators, there are a variety of decompositions of the total risk difference into mediation effects. We derive sharp and valid bounds for a number of mediation effects in the setting of two sequential mediators both with unmeasured confounding with the outcome. We provide five such bounds in the main text corresponding to two different decompositions of the total effect, as well as the controlled direct effect, with an additional thirty novel bounds provided in the supplementary materials corresponding to the terms of twenty-four four-way decompositions. We also show that, although it may seem that one can produce sharp bounds by adding or subtracting the limits of the sharp bounds for terms in a decomposition, this almost always produces valid, but not sharp bounds that can even be completely noninformative. We investigate the properties of the bounds by simulating random probability distributions under our causal model and illustrate how they are interpreted in a real data example. 
\end{abstract}

\noindent%
\vfill

\newpage
\spacingset{1.5} 

\clearpage

\section{Introduction}
In randomized trials with full compliance, the observed association between the intervention and the outcome has a causal interpretation as the total intervention effect through all possible pathways. Once a total intervention effect has been established, there is often additional interest in specific pathways and mechanisms through which the intervention may affect the outcome. For settings with a single binary mediator, \cite{robins1992identifiability} used counterfactual arguments to provide a formal framework for reasoning about direct effects and indirect (mediated) effects. \cite{pearl2001direct} proposed counterfactual definitions of direct and indirect effects for a single mediator. Specifically, \cite{pearl2001direct} distinguished between the controlled direct effect (CDE), which sets the mediator to a fixed value for each subject, and the natural direct effect (NDE), which sets the mediator to a  counterfactual value that may differ across subjects. Although the natural direct effect is more difficult to conceptualize, it has the appealing property that it adds up, together with the corresponding natural indirect effect (NIE), to the total effect. 

A problem with such effect decompositions is that the separate effect components are often not identified from data. Even when randomization rules out intervention-outcome confounding, there may still be unmeasured confounders for the mediator(s) and the outcome. If so, then any attempt to estimate the direct intervention effect by controlling for the mediator(s) will open back-door paths through the unmeasured confounders, thereby inducing an association between the intervention and the outcome \citep{robins1992identifiability}.

When causal effects are not identified, bounds for the target parameter of interest, i.e. a range that is guaranteed to include the true parameter value given the observed data, can be used to reduce the possible range of values that need to be considered. \citet{balke1994counterfactual} developed a method for deriving bounds for simple causal estimands based on linear programming techniques. For settings with a single binary mediator, \citet{cai2007non} used this linear programming technique to derive bounds on the CDE, assuming that the intervention is (or can be considered) randomized, but allowing for unmeasured confounding of the mediator and the outcome. \citet{sjolander2009bounds} provided analogous bounds for the NDE. \cite{kaufman2009analytic} provided bounds for both the CDE and the NDE while relaxing the assumption about the exposure being randomized. Other authors have derived bounds for the CDE and NDE under certain monotonicity assumptions \citep{vanderweele2011controlled,chiba2010bounds}, and for settings where the mediator has more than two levels \citep{miles2017partial}.

Recently, there has been a growing interest in effect decompositions with multiple mediators \cite[e.g][]{avin2005identifiability,albert2011generalized,vanderweele2014mediation,daniel2015causal,steen2017flexible}. However, this line of literature has focused on appropriate definitions and sufficient criteria for identification; to our knowledge, bounds for the nonidentified effect components in settings with more than one mediator have not been published. This is an important gap in the literature, since the introduction of multiple mediators poses identification problems that are not present in settings with a single mediator. Specifically, \cite{daniel2015causal} showed that when a mediator has a direct effect on a subsequent mediator, the indirect effect through the former mediator is not generally nonparametrically identified, even in the complete absence of unmeasured confounders. Hence, the importance of bounds is even stronger in setting with multiple mediators than in settings with a single mediator.  

We derive bounds for the setting of a two-armed randomized trial with two causally ordered binary mediators that are confounded with the binary outcome of interest. \cite{steen2017flexible} proposed a two-way and a three-way decomposition of the total effect, for which we provide bounds for each component. The two-way decomposition is obtained by separating the total effect into the direct effect of the intervention on the outcome and total indirect effects through both of the mediators or the `joint natural indirect effect', while the three-way decomposition further separates this joint natural indirect effect through the two mediators into two indirect effects. We also provide bounds for the controlled direct effects in this setting. In addition to providing sharp bounds for each term in the two-way and three-way decompositions, we demonstrate that in general the bounds for the separate terms of the decompositions cannot be combined to yield sharp bounds on their sum or difference. The exception is in the two-way decomposition when subtracting the limits of the joint indirect effect or the direct effect from the identified total effect, which we prove will always provide sharp bounds for the remaining effect. 

Further decompositions have been derived, and as is shown in \citet{daniel2015causal}, one of the indirect effects in the three-way decomposition can be further decomposed into the effect through the second mediator due to the effect of intervention directly on that mediator and the effect of the intervention through the first mediator on the second, resulting in a four-way decomposition. In addition to the bounds we provide in the main text, we provide bounds for the exhaustive list of the thirty-two terms appearing in the twenty-four possible decompositions of the total intervention effect into the four-way decompositions of \citet{daniel2015causal} in the supplementary materials. For each set of bounds we also provide easily downloadable \texttt{R} functions to facilitate their use. 

In practice, there is a tendency towards point estimation, even in the absence of a discussed or well defined estimand or even if that estimand is only identifiable under strong untestable assumptions. Although sensitivity analyses is sometimes used to mitigate concerns about assumptions, these procedures rarely provide an assumption free range of the true causal effect.  Due to the rising interest in the ``estimands framework'' for randomized clinical trials \citep{LipkovichRCT}, assumption free, i.e. nonparametric, bounds may become part of the standard for clinical trials analysis. In such settings, there are often multiple binary mediators, such as initiation of rescue mediation, withdrawal from treatment, or relapse or remission prior to death, making our bounds of practical relevance in these settings. Nonparametric bounds may sometimes be considered too wide to be useful in practice. In contrast, we believe that wide bounds are useful to present, since they highlight how little information the observed data contains about the target parameter, and how much a point estimate would have to rely on strong, potentially untestable assumptions.  

The paper is organized as follows. In Section \ref{sec:not} we provide our notation and outline our estimands and settings of interest. In Section \ref{results} we provide the bounds for each of the settings and estimands of interest. In Section \ref{sec:sims} we conduct some numerical studies to gain insight into the bounds we provide. In Section \ref{real} we illustrate our derived bounds in an illustrative data example from the \texttt{mediation} package in \texttt{R}. Code to reproduce all results in the simulations and real data example is available at \url{[blinded-url]} in addition to \texttt{R} functions to use the bounds in real data. Finally, in Section \ref{dis} we outline the limitations of our bounds and discussion future areas of research. 

\section{Preliminaries} \label{sec:not}
\subsection{Notation and Setting}
Let $X$ and $Y$ be the binary intervention and binary outcome of interest. Let $M_1$ and $M_2$ be two binary mediators on paths from $X$ to $Y$. Let $U$ be an unmeasured set of confounders between $Y$, $M_1$ and $M_2$ that are independent of $X$. The variables in this set have an unrestricted and unknown distribution, e.g. they may be a set of continuous and correlated unmeasured variables. We let $Y(X=x)$ denote the potential outcome $Y$ under the intervention which sets $X$ to $x$, and $Y(x, m_1, m_2)$ be the equivalent under an intervention which sets $X$ to $x$, $M_1$ to $m_1$, and $M_2$ to $m_2$.

\begin{figure}[h]
\centering

\begin{tikzpicture}
\node (E11) at (-2,0) {$X$};
\node (uh) at (3,0) {$U$};
\node (v1) at (-1,1) {$M_1$};
\node (v2) at (0,-1) {$M_2$};
\node (i) at (2,0) {$Y$};
\draw[-latex] (v1) -- (v2);
\draw[-latex] (v1) -- (i);
\draw[-latex] (v2) -- (i);
\draw[-latex] (E11) -- (v1);
\draw[-latex] (E11) -- (v2);
\draw[-latex] (uh) -- (v2);
\draw[-latex] (uh) -- (v1);
\draw[-latex] (uh) -- (i);
\draw[-latex] (E11) -- (i);
\end{tikzpicture}

\caption{Causal diagram of the setting of interest.  \label{b}}
\end{figure}
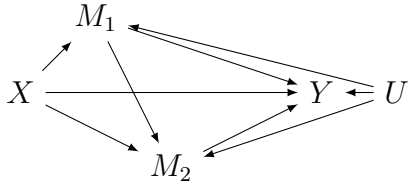

Our setting of interest is as depicted in the causal model or directed acyclic graph (DAG) in Figure \ref{b}. We interpret the DAG in Figure \ref{b} to represent a set of nonparametric structural equations such that: 
\begin{eqnarray*}
\label{seqs}
x &=& f_X(\epsilon_X)\\
m_1 &=& f_{M_1}(x, u, \epsilon_{m_{1}}) \\
m_2 &=& f_{M_2}(x, u, m_1, \epsilon_{m_{2}}) \\
y &=& f_Y(x, u, m_1, m_2, \epsilon_y) 
\end{eqnarray*}
for some response functions $f_X, f_{M_1}, f_{M_2}, f_Y$. The set of $\epsilon$'s represent `errors terms' due to omitted factors, which are assumed independent of $U$ and of each other. Given the values of the errors and the values of a variable's parents in the graph, the value of the variable is determined by its response function. The errors determine the manner in which the variable is determined from its parents. 

In this setting, the total effect (TE) is identified and can be estimated. However, none of the mediation effects are identified from the observed data due to the unmeasured confounder(s) $U$, which open pathways between $X$ and $Y$ when conditioning on $M_1$ or $M_2$. Both \citet{steen2017flexible} and \citet{daniel2015causal} provide rigorous proofs of this statement. 

Thus, in the following section we provide valid and sharp bounds for these estimands. These bounds are functions of the observed data distribution $p(Y,M_1,M_2|X)$. We say that the bounds are `sharp' when each value within the bounds is a possible value for the estimand, given the true probabilities that are estimable from data. Similarly, we say that the bounds are `valid' when no value outside the bounds is a possible value, given the true probabilities that are estimable from data.  

\subsection{Estimands}
There are several other potential outcomes to define in the setting with two mediators when the mediators are controlled `naturally'; what these estimands are depends on whether there is an effect of $M_1$ on $M_2$. When there is an effect of $M_1$ on $M_2$, as in our setting of interest, Figure \ref{b}, the potential outcome $Y(x, M_1(x_1), M_2(x_2,  M_1(x_3)))$ becomes relevant. This is the potential outcome of $Y$ under the interventions that sets $X$ to $x$, and $M_1$ to the value it would take on under the intervention that sets $X$ to $x_1$, and $M_2$ to the value it would take on under the intervention that sets $X$ to $x_2$ and $M_1$ to the value it would take on if $X$ was set to $x_3$. If there is no effect of $M_1$ on $M_2$, this potential outcome simplifies to $Y(X=x, M_1(x_1), M_2(x_2))$, as the value of $M_1$ will not impact $M_2$. In what follows, we focus on the setting of Figure \ref{b} where we allow an effect of $M_1$ on $M_2$.

The total effect of $X$ on $Y$ is defined as:
$$\mbox{TE} = p\{Y(1)=1\} -p\{Y(0)=1\}.$$
In the randomized and perfect compliance setting, such as our setting of interest, the TE is identified, and equal to $p(Y=1|X=1)-p(Y=1|X=0)$. We are also interested in the direct effect of $X$ on $Y$ holding the mediators to either a fixed level for all subjects, or the natural level they would have taken on had $X$ been set to $x$. Holding a single mediator to a fixed level gives what is referred to as the controlled direct effect. This can directly be extended to multiple mediators. We define the controlled direct effects (CDE) for two mediators, which have four possible levels defined by the values to which the mediators are being held, by: 
\begin{eqnarray*}
\mbox{CDE}\mbox{-}m_1 m_2 &=& p\{Y(1, M_1=m_1, M_2=m_2)=1\}\\ &-&p\{Y(0, M_1=m_1, M_2=m_2)=1\},
\end{eqnarray*}
$\mbox{ for } m_1, m_2 \in \{0, 1\}$.

This estimand fully describes the possible controlled direct effects in the setting of Figure \ref{b} below. When you hold the mediators to the counterfactual value they would have taken had you intervened on $X$, this is referred to as the natural direct effect, for a single mediator. Similar to the multiple mediators in the controlled direct effect case, we define the natural direct effects (NDE), in the setting of Figure \ref{b}. These are the estimands discussed in \citet{daniel2015causal}, and here we follow their nomenclature directly. 
\begin{eqnarray*}
\mbox{NDE}\mbox{-}x_1 x_2 x_3&=&p\{Y(1, M_1(x_1), M_2(x_2, M_1(x_3)))=1\}\\ 
&-&p\{Y(0, M_1(x_1), M_2(x_2, M_1(x_3)))=1\},
\end{eqnarray*}
for $x_1, x_2, x_3 \in \{0, 1\}$. The estimand $\mbox{NDE}\mbox{-}{000}$ was said to be the obvious extension of the pure natural direct effect in \citet{daniel2015causal}, and is the term in equation 7 in the decomposition of \citet{steen2017flexible}. 

We also consider the effect transmitted along either one or both mediators, the joint natural indirect effect: 
\begin{eqnarray*}
\mbox{JNIE}_{x}&=&p\{Y(x, M_1(1), M_2(1, M_1(1)))=1\}\\ &-& p\{Y(x, M_1(0), M_2(0, M_1(0)))=1\},
\end{eqnarray*}
for $x \in \{0, 1\}$. $\mbox{JNIE}_{1}$ is equal to the term in equation 6 of the decomposition as given in \citet{steen2017flexible}. 

We can further decompose this joint effect into two indirect effects, where the first component is the effect through $M_1$, directly and also through $M_2$, in the notation of \citet{daniel2015causal}: 
\begin{eqnarray*}
\mbox{MS}^2\mbox{-}\mbox{NIE}_{1}\mbox{-}x x_2 &=& p\{Y(x, M_1(1), M_2(x_2, M_1(1)))=1\}\\ &-& p\{Y(x, M_1(0), M_2(x_2, M_1(0)))=1\},
\end{eqnarray*}
for $x, x_2 \in \{0, 1\}$. $\mbox{MS}^2\mbox{-}\mbox{NIE}_{1}\mbox{-}11$ is equal to the term in equation 8 of the decomposition given in \citet{steen2017flexible}.

Finally, we consider the indirect effect through $M_2$, excluding the effect of $M_1$ through $M_2$, as given in \citet{daniel2015causal}: 
\begin{eqnarray*}
\mbox{NIE}_{2}\mbox{-}x x_1 x_3 &=&p\{Y(x, M_1(x_1), M_2(1, M_1(x_3)))=1\}\\ &-& p\{Y(x, M_1(x_1), M_2(0, M_1(x_3)))=1\},
\end{eqnarray*}
for $x, x_1, x_3 \in \{0, 1\}$. $\mbox{NIE}_{2}\mbox{-}100$ is equal to the term in equation 9 of the decomposition given in \citet{steen2017flexible}. Then the decomposition in \citet{steen2017flexible} is in our notation $$\mbox{TE} = \mbox{NDE}\mbox{-}{000} + \mbox{JNIE}_{1}$$ and is equivalent to
$$\mbox{TE}= \mbox{NDE}\mbox{-}{000} + \mbox{MS}^2\mbox{-}\mbox{NIE}_{1}\mbox{-}11 + \mbox{NIE}_{2}\mbox{-}100.$$ We focus on the bounds for the terms of these two decompositions in the main text. However, we also give the bounds on all terms in the twenty-four, four-term decompositions of \citet{daniel2015causal} in Section S3 of the supplementary materials and the decompositions are in Section S2. For this we need to define two more estimands, where again we use the same notation as \citet{daniel2015causal}. 

The indirect effect through $M_1$, excluding the effect of $M_1$ through $M_2$.
\begin{eqnarray*}
\mbox{NIE}_{1}\mbox{-}x x_2 x_3 &=&p\{Y(x, M_1(1), M_2(x_2, M_1(x_3)))=1\}\\ &-& p\{Y(x, M_1(0), M_2(x_2, M_1(x_3)))=1\},
\end{eqnarray*}
and the indirect effect of $M_1$ through $M_2$, excluding the direct effect of $M_1$.
\begin{eqnarray*}
\mbox{NIE}_{12}\mbox{-}x x_1 x_2 &=&p\{Y(x, M_1(x_1), M_2(x_2, M_1(1)))=1\}\\ &-& p\{Y(x, M_1(x_1), M_2(x_2, M_1(0)))=1\}.
\end{eqnarray*}
These terms add to the effect through $M_1$ and possibly also through $M_2$, $\mbox{MS}^2\mbox{-}\mbox{NIE}_{1}\mbox{-}11 =\mbox{NIE}_{1}\mbox{-}110 + \mbox{NIE}_{12}\mbox{-}111$, thus the four-way decomposition is given as in decomposition 3 of \citet{daniel2015causal} $$\mbox{TE} = \mbox{NDE}\mbox{-}{000}+ \mbox{NIE}_{1}\mbox{-}110 + \mbox{NIE}_{2}\mbox{-}100 +\mbox{NIE}_{12}\mbox{-}111.$$  

Define the short hand notation for the estimable probabilities as:
$$p_{ym_1m_2\cdot x} = p\{Y=y, M_1=m_1, M_2=m_2|X=x\}.$$ For example, $p_{111\cdot 1} = p\{Y=1, M_1=1, M_2=1|X=1\}$.

\section{Results} \label{results}

As the variables of interest in the DAG in Figure \ref{b} are all assumed binary, there exists a canonical partitioning of the unmeasured confounder $U$ into finite states, as described by \citet{balke1994counterfactual}. In this partitioning, the response function corresponding to each variable in the DAG is categorical with $2^{2^k}$ levels, where $k$ is the number of parents of that variable in the DAG. Ignoring the response function variable for $X$ since we condition on it, this leads to a total of $2^{2^1}\times 2^{2^2}\times 2^{2^3}=16,384$ probabilities associated with the joint response function variable distribution for $(M_1, M_2, Y)$, on which the thirty-two estimable probabilities form constraints that are used to bound the estimands of interest. The mediation effects of interest are the objectives that we maximize and minimize symbolically using vertex enumeration, resulting in bounds on the counterfactual probabilities in terms of the estimable data distribution. \\

\noindent \textbf{Result 1}:\\
The bounds given in \eqref{Beq:1} are valid and sharp for $\mbox{CDE}\mbox{-}m_1 m_2$ under Figure \ref{b}.

\begin{eqnarray}
\label{Beq:1}
-1 + p_{0m_1m_2.0} + p_{1m_1m_2.1} \leq \theta \leq  1 - p_{0m_1m_2.1} - p_{1m_1m_2.0}
\end{eqnarray}

With a small amount of algebra one can write the bounds in Result 1 in terms of the total effect, TE, with the lower bound as TE $- B(m_1,m_2)$, and the upper as TE $- B(m_1,m_2) + g(m_1,m_2)$. Here, 
$ B(m_1,m_2)$ is the sum of all differences, $p_{1m'_1m'_2.1} -p_{0m'_1m'_2.0}$ such that $m'_1m'_2 \neq m_1m_2$ and $ g(m_1,m_2) = p((M_1, M_2) \neq (m_1, m_2)|X = 0) + p((M_1, M_2) \neq (m_1, m_2)|X = 1)$. From this representation it is easy to see that when $g(m_1,m_2) = 0$, this also implies $B(m_1,m_2)=0$, and the bounds collapse to the single point, TE. Additionally, it can be seen from this that if TE $>B(m_1,m_2)$ then $\mbox{CDE}\mbox{-}m_1 m_2 >0.$ It is also immediately evident that when $p(M_2=m_2, M_1=m_1)=1$, then all NDE collapse to the CDE for the same $m_1,m_2$, and the same result applies. However, in this setting this is less mediation and more a deterministic relationship. 
\clearpage

\noindent \textbf{Result 2}:\\
The bounds given in \eqref{Beq:2} and \eqref{Beq:3} are valid and sharp for $\mbox{NDE}\mbox{-}000$  under Figure \ref{b}. 

\begin{align}
& \mbox{NDE}\mbox{-}000 \geq   \nonumber \\ 
\label{Beq:2}
& \max\left\{\begin{array}{l}
    p_{111\cdot 1} - p_{100\cdot 0} - p_{110\cdot 0} - p_{101\cdot 0} + p_{011\cdot 0} - 1\\
  -2 + p_{000\cdot 0} + p_{010\cdot 0} + 2*p_{001\cdot 0} + p_{101\cdot 0} + p_{101\cdot 1} + p_{011\cdot 0}\\
  -2 + p_{000\cdot 0} + 2*p_{010\cdot 0} + p_{110\cdot 0} + p_{110\cdot 1} + p_{001\cdot 0} + p_{011\cdot 0}\\
  -2 + 2*p_{000\cdot 0} + p_{100\cdot 0} + p_{100\cdot 1} + p_{010\cdot 0} + p_{001\cdot 0} + p_{011\cdot 0}\\
  -1 + p_{000\cdot 0} + p_{010\cdot 0} + p_{001\cdot 0} + p_{011\cdot 0}
\end{array}\right\},
\end{align}
\\
and
\begin{align}
&\mbox{NDE}\mbox{-}000 \leq \nonumber \\ 
\label{Beq:3}
&\min\left\{\begin{array}{l}
 1 + p_{000\cdot 0} - p_{010\cdot 1} - p_{110\cdot 0} + p_{001\cdot 0} + p_{011\cdot 0}\\
  1 + p_{000\cdot 0} + p_{010\cdot 0} - p_{001\cdot 1} - p_{101\cdot 0} + p_{011\cdot 0}\\
  1 + p_{000\cdot 0} - p_{111\cdot 0} + p_{010\cdot 0} + p_{001\cdot 0} - p_{011\cdot 1}\\ 
  p_{000\cdot 0} + p_{010\cdot 0} + p_{001\cdot 0} + p_{011\cdot 0}\\
  1 - p_{000\cdot 1} - p_{100\cdot 0} + p_{010\cdot 0} + p_{001\cdot 0} + p_{011\cdot 0}
\end{array}\right\}.
\end{align}

\clearpage
\noindent \textbf{Result 3}:\\
The bounds given in \eqref{Beq:4} and \eqref{Beq:5} are valid and sharp for $\mbox{JNIE}_{1}$  under Figure \ref{b}.

\begin{align}
&\mbox{JNIE}_{1}\geq  \nonumber \\ 
\label{Beq:4}
&\max\left\{\begin{array}{l}
  -1 + p_{111\cdot0} + p_{011\cdot 0} - p_{000\cdot 1} - p_{010\cdot 1} - p_{001\cdot 1} \\ 
  -p_{000\cdot 1} - p_{010\cdot 1} - p_{001\cdot 1} - p_{011\cdot 1}\\
  -1 - p_{000\cdot 1} - p_{010\cdot 1} + p_{001\cdot 0} + p_{101\cdot 0} - p_{011\cdot 1}\\
  -1 - p_{000\cdot 1} + p_{010\cdot 0} + p_{110\cdot 0} - p_{001\cdot 1} - p_{011\cdot 1}\\
  -1 + p_{000\cdot 0} + p_{100\cdot 0} - p_{010\cdot 1} - p_{001\cdot 1} - p_{011\cdot 1}
\end{array}\right\},
\end{align}
\\
and
\begin{align}
&\mbox{JNIE}_{1} \leq \nonumber \\ 
\label{Beq:5}
&\min\left\{\begin{array}{l}
 2 - p_{000\cdot 0} - p_{000\cdot 1} - p_{100\cdot 0} - p_{100\cdot 1} - p_{010\cdot 1} - p_{001\cdot 1} - p_{011\cdot 1}\\
  1 - p_{000\cdot 1} - p_{010\cdot 1} - p_{001\cdot 1} - p_{011\cdot 1}\\
  2 - p_{000\cdot 1} - p_{010\cdot 0} - p_{010\cdot 1} - p_{110\cdot 0} - p_{110\cdot 1} - p_{001\cdot 1} - p_{011\cdot 1}\\
  2 - p_{000\cdot 1} - p_{010\cdot 1} - p_{001\cdot 0} - p_{001\cdot 1} - p_{101\cdot 0} - p_{101\cdot 1} - p_{011\cdot 1}\\
  1 - p_{111\cdot 0} - p_{011\cdot 0} + p_{100\cdot 1} + p_{110\cdot 1} + p_{101\cdot 1} 
\end{array}\right\}.
\end{align}
\clearpage

\noindent \textbf{Result 4}:\\
The bounds given in \eqref{Beq:6} and \eqref{Beq:7} are valid and sharp for $\mbox{MS}^2\mbox{-}\mbox{NIE}_{1}\mbox{-}11$ under Figure \ref{b}.

\begin{align}
&\mbox{MS}^2\mbox{-}\mbox{NIE}_{1}\mbox{-}11 \geq  \nonumber \\ 
\label{Beq:6}
&\max\left\{\begin{array}{l}
 -p_{000\cdot 0} - p_{000\cdot 1} - p_{100\cdot 0} - p_{001\cdot 0} - p_{001\cdot 1} - p_{101\cdot 0}\\
  -p_{000\cdot 1} - p_{010\cdot 1} - p_{001\cdot 1} - p_{011\cdot 1}\\
  -1 + p_{000\cdot 0} + p_{100\cdot 0} - p_{010\cdot 1} + p_{001\cdot 0} + p_{101\cdot 0} - p_{011\cdot 1}
\end{array}\right\},
\end{align}
\\
and
\begin{align}
&\mbox{MS}^2\mbox{-}\mbox{NIE}_{1}\mbox{-}11 \leq \nonumber \\ 
\label{Beq:7}
&\min\left\{\begin{array}{l}
1 - p_{000\cdot 0} - p_{100\cdot 0} - p_{001\cdot 0} - p_{101\cdot 0} + p_{111 \cdot1} + p_{110 \cdot1}\\ 
  1 - p_{000\cdot 1} - p_{010\cdot 1} - p_{001\cdot 1} - p_{011\cdot 1}\\
  p_{000\cdot 0} + p_{100\cdot 0} + p_{100\cdot 1} + p_{001\cdot 0} + p_{101\cdot 0} + p_{101\cdot 1}
\end{array}\right\}.
\end{align}
\clearpage

\noindent \textbf{Result 5}:\\
The bounds given in \eqref{Beq:8} and \eqref{Beq:9} are valid and sharp for $\mbox{NIE}_{2}\mbox{-}100$ under Figure \ref{b}.

\begin{align}
&\mbox{NIE}_{2}\mbox{-}100 \geq  \nonumber \\ 
\label{Beq:8}
&\max\left\{\begin{array}{l}
  -1 + p_{011\cdot 0} + p_{111\cdot 0} - p_{000\cdot 1}  - p_{100\cdot 1} - p_{010\cdot 1} - p_{001\cdot 1} - p_{101\cdot 1} \\ 
  -1 + p_{110\cdot 1} + p_{111\cdot 1} - p_{000\cdot 0} - p_{100\cdot 0} - p_{001\cdot 0} - p_{101\cdot 0} \\ 
  -2 + p_{100\cdot 1} + p_{001\cdot 0} + p_{001\cdot 1} + p_{101\cdot 0} + p_{101\cdot 1} \\
  -2 + p_{000\cdot 0} + p_{100\cdot 0} + p_{100\cdot 1} + p_{001\cdot 0} + p_{101\cdot 0} + p_{101\cdot 1} \\
  -1 - p_{000\cdot 1} - p_{100\cdot 1} + p_{010\cdot 0} + p_{110\cdot 0} - p_{001\cdot 1} - p_{101\cdot 1} - p_{011\cdot 1} \\
  -2 + p_{000\cdot 0} + p_{000\cdot 1} + p_{100\cdot 0} + p_{100\cdot 1} + p_{101\cdot 1} \\
  -1
\end{array}\right\},
\end{align}
\\
and
\begin{align}
&\mbox{NIE}_{2}\mbox{-}100 \leq \nonumber \\ 
\label{Beq:9}
&\min\left\{\begin{array}{l}
 2 - p_{000\cdot 0} - p_{000\cdot 1} - p_{100\cdot 0} - p_{100\cdot 1} - p_{001\cdot 1}\\
  2 - p_{000\cdot 0} - p_{000\cdot 1} - p_{100\cdot 0} - p_{001\cdot 0} - p_{001\cdot 1} - p_{101\cdot 0} \\
  2 - p_{000\cdot 1} - p_{001\cdot 0} - p_{001\cdot 1} - p_{101\cdot 0} - p_{101\cdot 1} \\
  2 - p_{010\cdot 0} - p_{010\cdot 1} - p_{110\cdot 0} - p_{110\cdot 1} - p_{011\cdot 1} \\
  1 + p_{000\cdot 0} + p_{100\cdot 0} - p_{010\cdot 1} + p_{001\cdot 0} + p_{101\cdot 0} - p_{011\cdot 1} \\
  1 - p_{011\cdot 0} - p_{111\cdot 0} + p_{000\cdot 1} + p_{100\cdot 1} + p_{110\cdot 1} + p_{001\cdot 1} + p_{101\cdot 1} \\ 
  1
\end{array}\right\}.
\end{align}
\bigskip

To compare the bounds presented above, it is useful to consider alternative valid bounds defined by `subtraction procedures' or `addition procedures', as follows. Consider an estimand of interest $\theta$ such that it can be decomposed into several terms $\theta = \sum_{i=1}^I \gamma_i.$ Let $(l_i, u_i)$ be the valid and sharp bounds for each  $\gamma_i$, respectively. Given the decomposition of $\theta$ we have $\gamma_i = \theta - \sum_{j\neq i} \gamma_j$ so that alternative bounds for $\gamma_i$ are given by $(l_i^*, u_i^*)$, where $l_i^* = \theta_i - \sum_{j\neq i} u_j$, and $u_i^*= \theta_i - \sum_{j\neq i} l_j$. This is what we will call the subtraction procedure. Similarly, one can obtain valid bounds for $\theta$ by adding the bounds for $\gamma_i$ $u^*=\sum_i u_i$, and $l^*= \sum_i l_i$. This is what we will call the addition procedure. These procedures will not always produce sharp bounds, as is demonstrated in the real data example, but there are at least some special cases where the subtraction procedure can be used to produce valid and sharp bounds.

\noindent \textbf{Result 6}:\\
For any two-way decomposition of the identified TE the subtraction of the limits of sharp bounds for one term in the decomposition from the TE will produce the sharp bounds for the other term.\\

Result 6 is easily proven by considering three estimands, $\theta$, $\gamma_1$ and $\gamma_2$, such that $\theta = \gamma_1+\gamma_2$ where $\theta$ is identified, but $\gamma_1$ and $\gamma_2$ are not. Suppose that $(l_1,u_1)$ and $(l_2,u_2)$ are valid and sharp bounds for $\gamma_1$ and $\gamma_2$, respectively. We know that $\gamma_2 = \theta - \gamma_1$ and that the validity of the bounds for $\gamma_1$ implies that $\theta- u_1 \leq \gamma_2 \leq \theta - l_1$ is valid. 
In addition, the sharpness of the bounds for $\gamma_1$ imply that any point in $\theta- u_1 \leq \gamma_2 \leq \theta - l_1$  is possible, which then implies that $\theta - u_1 = l_2$ and $\theta- l_1 = u_2$, if the bounds for $\gamma_2$ are also valid and sharp. 

This is also easy to see by looking at the first panel of Figure \ref{fig:my_label4}. If the bounds are sharp, then the triangles must be of the same size and mirror images of themselves over the line defined by the TE. If one of the triangles is smaller on one side, then by projecting across the TE line you could produce narrower bounds for the other estimand. This contradicts the conditions of the result. Alternatively, if one of the triangles is too large, projecting from the smaller triangle towards the term you are subtracting results in the same contradiction, thus proving the result. 

Result 6 does not generalize to a three-way decomposition. For example, $\theta = \gamma_1+\gamma_2 +\gamma_3$, where again $\theta$ is identified, but $\gamma_1$, $\gamma_2$ and $\gamma_3$ are not. We now also have sharp and valid bounds for $\gamma_3$ $(l_3,u_3)$. We can again bound $\gamma_2$ by $\theta - u_1 - u_3 \leq \gamma_2 \leq \theta - l_1 - l_3$. By the validity of the bounds for $\gamma_1$ and $\gamma_3$ this bound is valid for $\gamma_2$. However, they are only sharp if the limiting values $l_1$ and $l_3$ are simultaneously possible values of $\gamma_1$ and $\gamma_3$, and $u_1$ and $u_3$ are also simultaneously possible. If the bounds are not constructed by considering the $\gamma$'s jointly, there is no guarantee that $\gamma_1$ can take on the value $l_1$, while $\gamma_3$ is equal to $l_3$, or  $\gamma_3$ taken on the value $u_3$ while  $\gamma_1$ is equal to $u_1$. Even in the case of a two-way decomposition it is clear that $l_1+l_2$ and $u_1+u_2$ will not always provide sharp bounds for the identified $\theta$, unless $l_1=u_1$ and $l_2=u_2$. Additionally, as is demonstrated in our real data example, if $\theta$ is also not identified with $\theta=\gamma_1+\gamma_2$ and has valid and sharp bounds $(l_0,u_0)$, the bounds derived for $\gamma_2$ using the subtraction procedure, $l_0-u_1 \leq \gamma_2 \leq u_0-l_1$, have no guarantee of being sharp or even informative, i.e. not containing -1 and 1. 

This makes it clear that the bounds for $\mbox{JNIE}_{1}$ are generally narrower than those resulting from adding the limits of the bounds for each term in its decomposition; this is demonstrated in the lower panel of Figure \ref{fig:my_label4}. In this example, the bounds for $\mbox{NIE}_{2}\mbox{-}100$ are wider than those for $\mbox{JNIE}_{1}$, making clear that the addition procedure will not produce sharp bounds. 
Although in our real data example bounds for $\mbox{MS}^2\mbox{-}\mbox{NIE}_{1}\mbox{-}11$ are  narrower than $\mbox{JNIE}_{1}$, this will not always be the case. The bounds for $\mbox{MS}^2\mbox{-}\mbox{NIE}_{1}\mbox{-}11$ differ by a single term in each of the upper and lower bounds \eqref{Beq:4} and \eqref{Beq:5} for $\mbox{JNIE}_{1}$. These terms, when active, tend to make the bounds narrower, although this does not often mean they exclude zero, as we demonstrate in simulations below.  

\section{Simulations} \label{sec:sims}
We randomly generated counterfactual probabilities and then use the constraints implied by the DAG to generate the true estimable probabilities, allowing us to ensure that we are generating distributions under the DAG. We generated the K = 16,384 dimensional counterfactual probability distribution vector by sampling from a Dirichlet distribution with the vector of parameters $\overline{\alpha}=\{\alpha_1,\ldots, \alpha_K\}$, that has probability distribution function:

$$f(q_1,\ldots,q_K;\alpha_1,\ldots,\alpha_K)=\frac{1}{B(\overline{\alpha})}\sum^{K}_{i=1} q_i^{\alpha_i-1},$$
where $B(\cdot)$ is the multivariate beta function, and where $\sum_i q_i = 1$. 

The generated counterfactuals describe a 16,384-dimensional space that is difficult to explore in any exhaustive manner. We consider two special cases, points at the vertices of the space where a single counterfactual probability is 1 and all others are zero, and the symmetric Dirichlet distribution, where all $\alpha_i$ are equal. Additionally we consider only two characteristics of the bounds under each of these special cases, the width, with anything less than 2 providing information over the always valid (-1,1) bounds for any risk difference, and if the bounds cover zero, thus not providing evidence against the causal null hypothesis. 

\begin{table}[ht]
\caption{\label{tab01} Counts (proportion) out of the 16,384 vertices where the lower bound equals the label in the rows, and the upper bound equals the label in the columns. }
\centering
\begin{tabular}{lc|p{14mm}p{14mm}p{14mm}|cc|p{14mm}p{14mm}}
& \multicolumn{4}{c|}{$\mbox{NDE}$-000} & \multicolumn{4}{c}{$\mbox{JNIE}_{1}$} \\ 
& & \multicolumn{3}{c|}{Upper bound} & & & \multicolumn{2}{c}{Upper bound} \\
& & -1 & 0 & 1 & & & 0 & 1 \\
 \hline
\multirow{3}{*}{\rotatebox[origin=c]{90}{Lower bound}} &-1 & 1024 (0.0625) & 6144 (0.375) & 0 & & -1 & 6144 (0.375) & 0 \\
&0 & 0 & 2048 (0.125) & 6144 (0.375) & & 0 & 4096 (0.25) & 6144 (0.375) \\
&1 & 0 & 0 & 1024 (0.0625) & & & & \\
\cline{2-9}
&\multicolumn{4}{c|}{$\mbox{MS}^2\mbox{-}\mbox{NIE}_{1}\mbox{-}11$} & \multicolumn{4}{c}{$\mbox{NIE}_{2}\mbox{-}100$} \\ 
\multirow{3}{*}{\rotatebox[origin=c]{90}{Lower bound}} & & & 0 & 1 & & & 0 & 1 \\
 \cline{2-9}
 &-1 & & 4096 (0.25) & 0 & & -1 & 2048 (0.125) & 8192 (0.50) \\ 
&0 & & 8192 (0.50) & 4096 (0.25) & & 0 & 4096 (0.25) & 2048 (0.125) \\
\end{tabular}
\end{table}

When considering the vertices of the 16,384-dimensional space, we found that the width of the bounds under these distributions is bimodal/trimodal, see Table \ref{tab01}. Specifically, for the $\mbox{NDE}$-000 and $\mbox{JNIE}_{1}$, the widths of the bounds are 0 for 25\% of the vertices and 1 for 75\% of the vertices. For the $\mbox{MS}^2\mbox{-}\mbox{NIE}_{1}\mbox{-}11$, the widths of the bounds are 0 for half of the vertices and 1 for the other half. In all those cases, all of these bounds either exclude 0 or 0 is the lower or upper limit of the bounds. For the $\mbox{NIE}_{2}\mbox{-}100$, the widths of the bounds are 0 for 25\%, 1 for 25\%, and 2 for 50\% of the vertices. In the cases where the widths of the bounds are not 2, they either exclude 0 or 0 is the lower or upper limit of the bounds.

\begin{figure}
    \centering
    \includegraphics[width=0.98\textwidth]{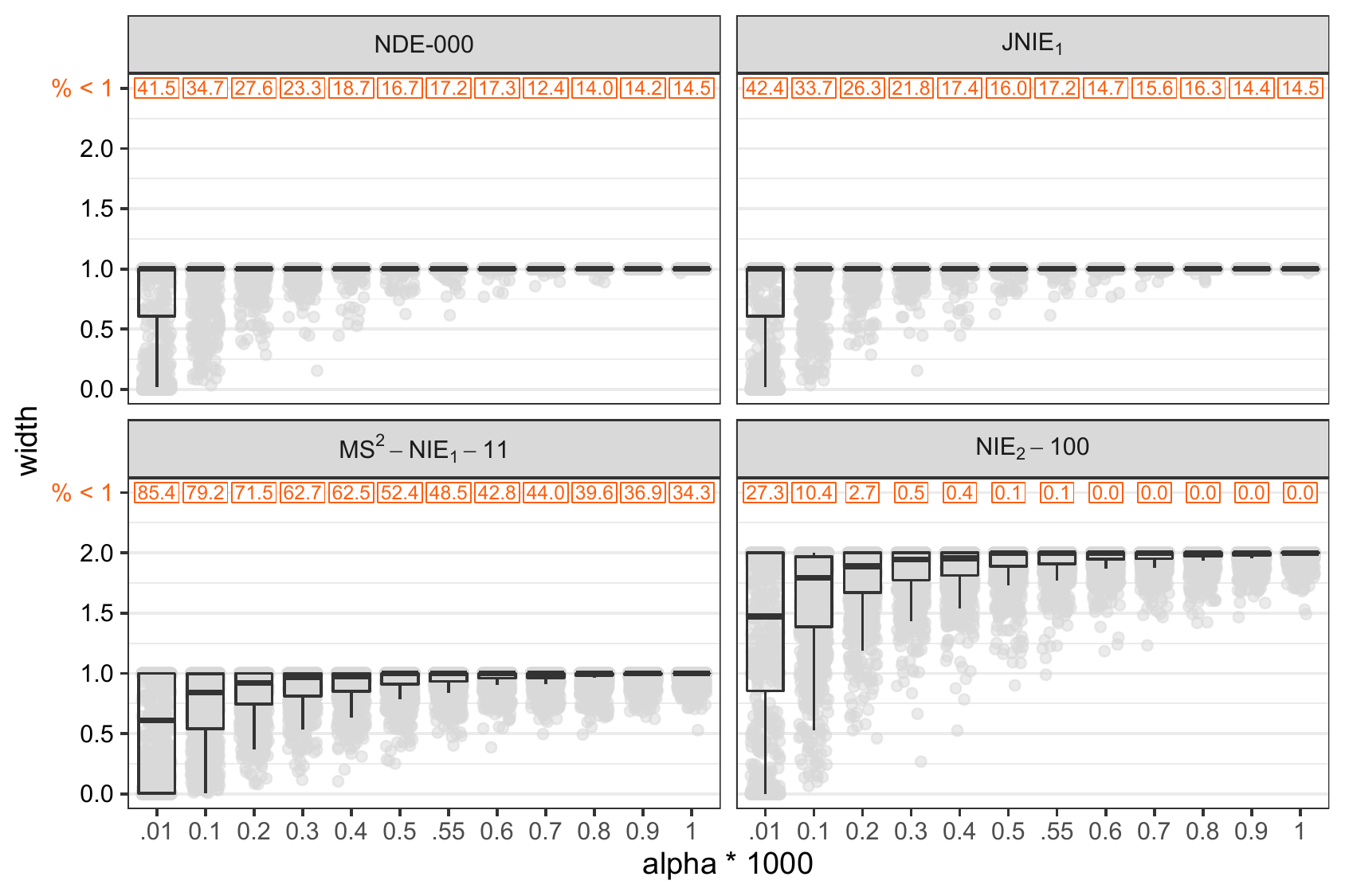}
    \caption{Width of bounds (points) with boxplots for 1,000 simulated distributions generated under different values of the parameter of a symmetric Dirichlet distribution. The orange numbers at the top of each figure indicate the \% of bounds that have width less than 1. For all estimands other than NIE$_2$-100, 100 minus this \% provides the bounds with width of exactly 1, as the bounds for these estimands never exceeds 1. For NIE$_2$-100 it can be seen that the bounds width is often larger than 1 and in many cases is width 2.}
    \label{fig:my_label1}
\end{figure}

\begin{figure}
    \centering
    \includegraphics[scale=.9]{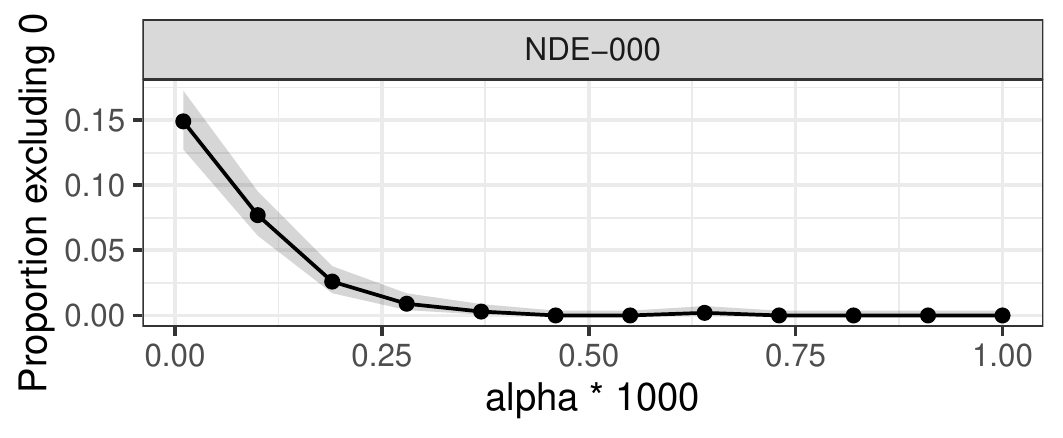}
    \caption{Proportion of bounds for the NDE-000 that exclude 0 out of 1,000 simulated distributions generated under different values of the parameter of a symmetric Dirichlet distribution. Grey ribbon indicates Clopper-Pearson 95\% confidence limits for the proportion.}
    \label{fig:my_label2}
\end{figure}

When the $\alpha_i$ take a common value decreasing from 1 to 0, the counterfactual space has more and more of a bowl shape, such that the areas where all the counterfactual probabilities would be non-zero and equal has the lowest probability of being generated and the vertices where only one probability is non-zero has the highest. Figures \ref{fig:my_label1} and \ref{fig:my_label2} trace the path from all the $\alpha_i$ being nearly zero (0.000001) to all being equal to 0.001, for the width and exclusion of 0, respectively. As can be seen in the figures, over the 1,000 simulations at each level of the $\alpha_i$ only at very low values do all of the estimands have a large proportion of bounds with widths less than 1. However, at no value of the $\alpha_i$ do the widths exceed 1 for the direct effect, the joint indirect effect, or the indirect effect through $M_1$. The bounds for the effect through $M_2$ only are quite wide, often being greater than 1, with a large proportion having a width of 2 at all $\alpha_i=0.001$. Additionally, we see that in most cases the bounds contain zero for the indirect effects, even when, in the case of the MS$^2$-NIE$_1$-111, there are a large proportion of the bounds that have width less than 1. The bounds for NDE-000, on the other hand, often exclude zero for low $\alpha_i$ values. This suggests that when many of the counterfactual probabilities are zero, the bounds are more informative.  This is not surprising, as counterfactual probabilities being zero can be considered constraints, such as the well-known constraint in the instrumental variable setting called no defiers or monotonicity. In a 16,384-dimensional space, such constraints are more difficult to give intuitive names. When the $\alpha_i$ are equal and held to be 1 this is the same as putting a flat uniform distribution on the probabilities. The trend in Figure \ref{fig:my_label1} continues, with estimated probability of the width of the bounds being less than one approaching zero as the $\alpha_i$ approaches one.

Our simulations do not give a complete picture of what is occurring in the full 16,384-dimensional space, as we have also found examples of sections of the space where two counterfactuals are both nonzero and the rest are equal but nonzero, where both the width of the bounds are less than 1 and the bounds exclude zero for the NDE-000, JNIE$_1$ and $\mbox{MS}^2\mbox{-}\mbox{NIE}_{1}\mbox{-}11$ effects. Given our limited exploration we have not uncovered a discernible pattern, other than the bounds tend to be more informative when a large number of the counterfactuals are zero. 

\section{Real Data Example} \label{real}
\begin{figure}[h]
    \centering
    \includegraphics[width = .95\textwidth]{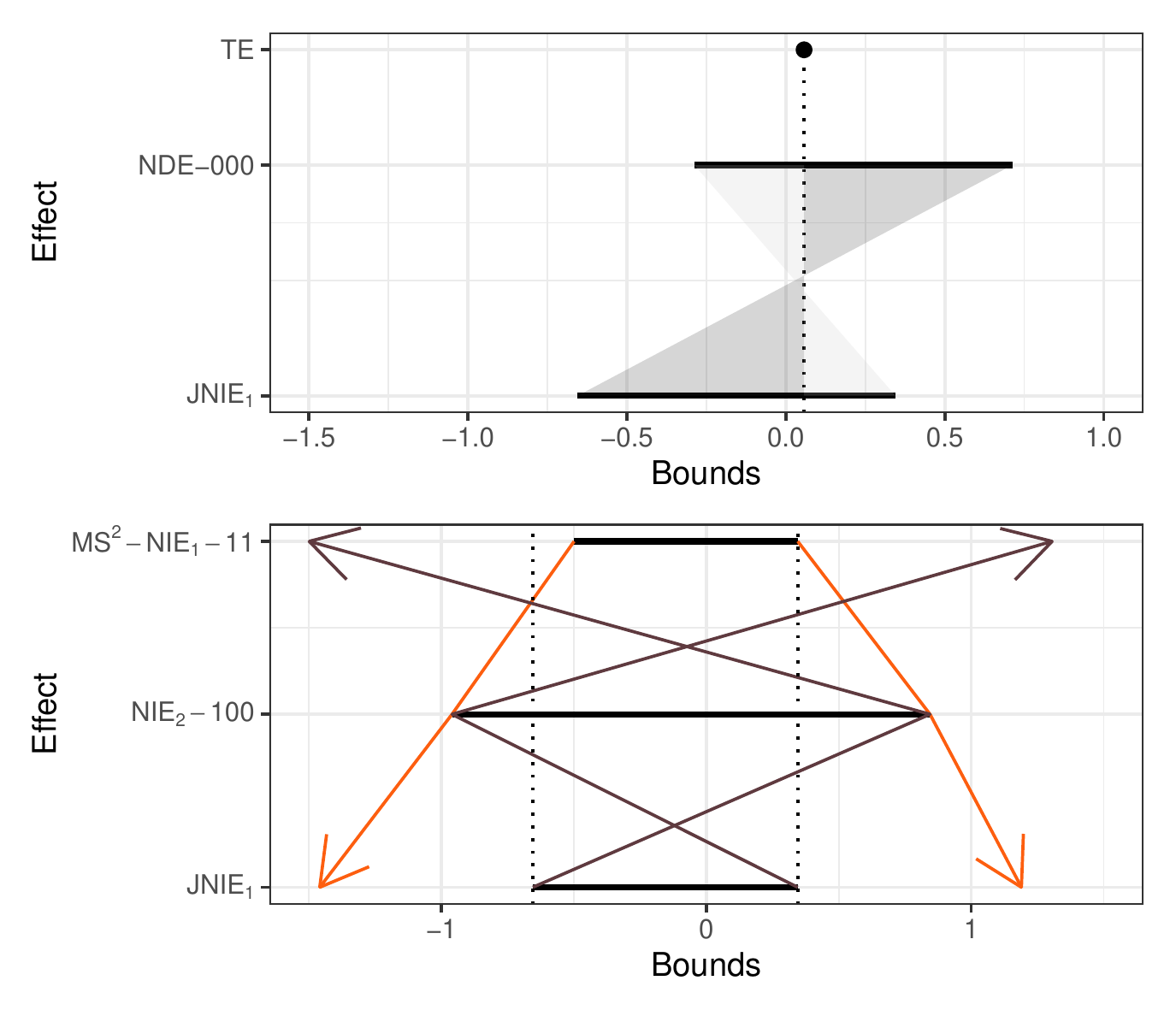}
    \caption{Bounds for the jobs dataset. The orange arrows show the addition procedure, summing the upper and lower bounds values of  MSE$^2$-NIE$_1$-11 and NIE$_2$-100 to obtain valid, but not sharp bounds for NIE$_1$. The brown arrows demonstrate the subtraction procedure where an alternative lower bound for MSE$^2$-NIE$_1$-11 is derived by subtracting the upper bound value of NIE$_2$-100 from the lower bound value of JNIE$_1$, and an alternative upper bound for MSE$^2$-NIE$_1$-11 is derived by subtracting the lower bound value of NIE$_2$-100 from the upper bound value of JNIE$_1$. Again, this is valid but not sharp. \label{fig:my_label4}}
\end{figure}

We illustrate our bounds using the \texttt{jobs} dataset from the \texttt{mediation} package in R \citep{mediation}. These data come from a randomized experiment designed to investigate the efficacy of a job training intervention on unemployed workers. In the experiment, 899 eligible unemployed workers completed a pre-screening questionnaire and were then randomly assigned to treatment, which consisted of participation in job skills workshops, or control, who received a booklet of job search tips. The randomization was 2:1 in favor of treatment with 600 assigned to treatment and 299 to control. The primary outcome is a binary variable representing whether the respondent had become employed, with 35\% on the treatment arm  and 29\% on the control arm becoming employed at the end of the study. The mediators of interest are a binary indicator of depressive symptoms after treatment, and a binary indicator of high job seeking self-efficacy. We believe that depression is the first mediator, or $M_1$, which may have a causal effect on job seeking self-efficacy, $M_2$. The bounds are displayed in Figure \ref{fig:my_label4}. 

As can be seen in the first panel of Figure \ref{fig:my_label4}, the total effect was estimated to be barely above zero, $0.057$, which by an exact test is not significantly different than 0, with a p-value of 0.10 and 95\% CI $(-0.01, 0.12)$. Also in this panel it can be seen that the bounds for the natural direct effect $(-0.29, 0.71)$ and the joint indirect effect $(-0.65, 0.34)$ are mirror images across the line of the total effect. This is as expected, given Result 6. Looking at the second panel in Figure \ref{fig:my_label4} it can be seen that the bounds for the MS$^2$-NIE$_1$-111 $(-0.50, 0.34)$ are narrower than both the natural indirect effect through high job seeking $(-0.96, 0.84)$ and the joint indirect effect. It can also be seen that the addition procedure used on the bounds for MS$^2$-NIE$_1$-111 and NIE$_2$-100 is wider than the sharp bounds for JNIE$_1$, as the bounds for NIE$_2$-100 alone are wider than those for JNIE$_1$. The subtraction procedure of the lower bound of NIE$_2$-100 from the upper bound of the JNIE$_1$, as well as the upper from the lower, produce much wider bounds for MS$^2$-NIE$_1$-111. This is, again, as expected as the bounds are sharp. In this case, the addition procedure and subtraction procedure both produce noninformative bounds, i.e. the bounds are outside the possible range of the causal effect $(-1,1)$. Additionally, although it was not the focus of this paper, Result 1 provides the bounds for the controlled direct effects and they are as follows in these data: CDE-00 $\in (-0.71, 0.79)$, CDE-01 $\in (-0.51, 0.54)$, CDE-10  $\in (-0.84, 0.86)$, and CDE-11 $\in (-0.87, 0.87)$. Each CDE bound covers zero and has a width greater than one, with CDE-01 having the narrowest width of 1.03. 

\section{Discussion} \label{dis}
We present bounds for estimands of mediation effects in the two-mediator setting, including many of the estimands considered in \citet{daniel2015causal, steen2017flexible}. We show that in many cases the bounds will be narrower than one, and this occurs frequently when a large number of the counterfactual probabilities are zero. We also find that the addition or subtraction of sharp bounds for two estimands that are not identified does not produce, in general, sharp bounds for the estimand of their addition or subtraction. Additionally, the subtraction of sharp bounds does not produce sharp bounds for the remaining term(s) unless one of the estimands in the subtraction is identified. We prove that in a two-way decomposition of the identified TE, the subtraction of a set of sharp bounds will always produce sharp bounds for the other term. 

\cite{daniel2015causal} showed that many of the mediation estimands are not identified even in the complete absence of unmeasured confounders, making bounds of particular relevance. However, it should be noted that the bounds we provide are not generated under the assumption of no confounding. So, while the bounds will be valid under the assumptions of no confounding, one could possibly obtain narrower sharp bounds by including that assumption in the derivation.

Although we explore two specific scenarios in the simulations, exploring the full space, perhaps with some type of greedy algorithm, to determine if there is a pattern which corresponds to constraints is beyond the scope of this paper, but an area of future research for the authors. Additionally, in exploring the space in greater detail, we will likely find constraints to consider in the construction of bounds, which may result in narrower or different sets of bounds for these decomposition terms. 

We provide a large number of  bounds, and insight into their interpretation, but we do not discuss an estimation procedure. As the terms of the bounds are all linear combinations of conditional probabilities, estimation is straightforward. For inference, the same bootstrap procedure suggested in \citet{Gabriel2020} and \citet{horowitz2000nonparametric} can be used here. In both of those papers, extensive empirical results were provided showing nominal coverage of quantile bootstrap confidence intervals for similarly constructed bounds. Although we believe that the bootstrap is likely theoretically justified, there may be closed form variance estimates for many, if not all, of the bounds. We do not focus on accounting for the sampling variability here, as for any reasonable sample size, we expect the uncertainty in causal effects due to unmeasured confounding to be far greater than uncertainty due to sampling variability. 

Although we only consider the setting where there are unmeasured and no measured confounders, in the simplest case where the measured confounders are discrete and bounded and do not restrict the impact of the unmeasured confounders on the mediators and the outcome, the bounds can be applied within levels of the measured confounders. More complex settings would need to be considered on a case-by-case basis and may result in narrower sharp bounds under particular assumptions. There are other scenarios where the same linear programming method may be used to determine if the bounds change, such as allowing the intervention to have a direct effect on the confounder. Defining the general conditions under which a DAG and target define a linear programming problem, and the automated checking of these conditions is an area of future research for the authors. Finally, our present results are limited to binary exposures, mediators, and outcomes so it would be worthwhile to extend these to multicategorical variables.

\bibliographystyle{plainnat}
\bibliography{main}
\end{document}